\documentclass{sf2a-conf2015}
\usepackage{graphicx}
\usepackage{hyperref}
\usepackage[]{natbib}  
\usepackage{epstopdf}
\usepackage{amsmath}

\def\BibTeX{{\rm B\kern-.05em{\sc i\kern-.025em b}\kern-.08em
    T\kern-.1667em\lower.7ex\hbox{E}\kern-.125emX}}
\bibpunct{(}{)}{;}{a}{}{,}  
\newcommand{\kepler}{\textit{Kepler}}
\newcommand{\q}[1]{``#1''}
\newcommand{\thetta}{\ensuremath{\alpha \text{, } \beta \text{, and } \sigma}}

\begin{document}

\TitreGlobal{SF2A 2015}


\title{Transit-Depth Metallicity Correlation: A Bayesian Approach}

\runningtitle{Transit-Depth Metallicity Correlation}

\author{P. Sarkis}\address{Department of Physics \& Astronomy, Notre Dame University, Lebanon \\
\href{mailto:pjsarkis01@ndu.edu.lb}{pjsarkis01@ndu.edu.lb}}
\author{C. Nehm\'e$^{1,}$}\address{CEA Saclay, IRFU, France}

\setcounter{page}{237}


\maketitle


\begin{abstract}
A negative correlation was previously reported between the transit depth of \textit{Kepler}'s 
Q1-Q12 gas giant candidates and the stellar metallicity. In this present work, we revisit this 
correlation
to better understand the role of the stellar metallicity in the formation of giant planets,
in particular, to investigate the effect of the metallicity on the transit depth.
We selected the 82 confirmed giant planets from the cumulative catalogue.
This is the first large and homogenous sample of confirmed giant planets 
used to study this correlation. Such samples are suitable to perform robust statistical analysis. 
We present the first hierarchical Bayesian linear regression model to revise this correlation. The 
advantages of using a Bayesian framework are to incorporate measurement errors in the model and 
to quantify both the intrinsic scatter and the uncertainties on the parameters of the model.
Our statistical analysis reveals no correlation between the transit depth of confirmed 
giant planets and the stellar metallicity.
\end{abstract}



\section{Introduction}

NASA's \kepler\ mission has revolutionized 
the field of extrasolar planets and now, more than ever, it 
is possible to put statistical constraints on 
the observed planet properties and on the theories of planet formation. 
Thousands of exoplanets have been discovered by \kepler\
which allows one to look at the planets as a population and not 
as single planets. 
Clues on the nature of giant planet formation might be revealed from 
observational trends. With such large and homogenous samples, it is now
possible to compare observed correlations to theories.
\citet[hereafter DR12]{Dodson:2012} reported a negative correlation between the transit depth 
of \kepler's gas giant planets and the metallicity of the host star. 
DR12's sample consisted of 213 giant candidates with estimated radii 
of $5-20R_\oplus$. The author argued that her sample may be contaminated by false positives
and interpreted the result as evidence that metal-rich planets of a
given mass are denser than their metal-poor counterparts, leading to smaller radii
\citep{Fortney:2010}. DR12 did not include in her statistical analysis the 
uncertainties on the  transit depth and on the stellar metallicity although 
they are very important and could bias the result.
In this present work and in order not to contaminate our sample with 
false positives, we revise the transit depth - metallicity 
correlation for all the confirmed giant planets detected by \kepler.
Moreover, we use Bayesian statistics to incorporate the 
measurement uncertainties in our analysis.

\section{Sample Selection}

For this study we used the cumulative catalog of planets detected by the NASA
\kepler\ mission which, as of April 2015, consisted of the 
latest Q1-Q16 catalog \citep{Mullally:2015}.
Following \cite{Dodson:2012}, we define gas giant planets as planets 
that have a radius between $5R_\oplus < R_p < 20R_\oplus$. We selected only the 
planets with a SNR $ > 7.1$ to avoid KOIs (or \kepler\ Objects of Interest)
with noisy lightcurves. 
The stellar parameters were taken from the \kepler\ stellar Q1-Q16 database 
\citep{Huber:2014}.
We ended up with a sample of 373 giant planets
of which 82 are confirmed gas giant exoplanets.
Note that the stellar and the planetary parameters provided
by \kepler's catalog have asymmetric upper and lower uncertainties. To get the 1$\sigma$
error bar we calculated their average.

\section{The Method: Hierarchical Bayesian Modeling}

\begin{figure}[t!]
\centering
\includegraphics[width=0.6\textwidth,clip]{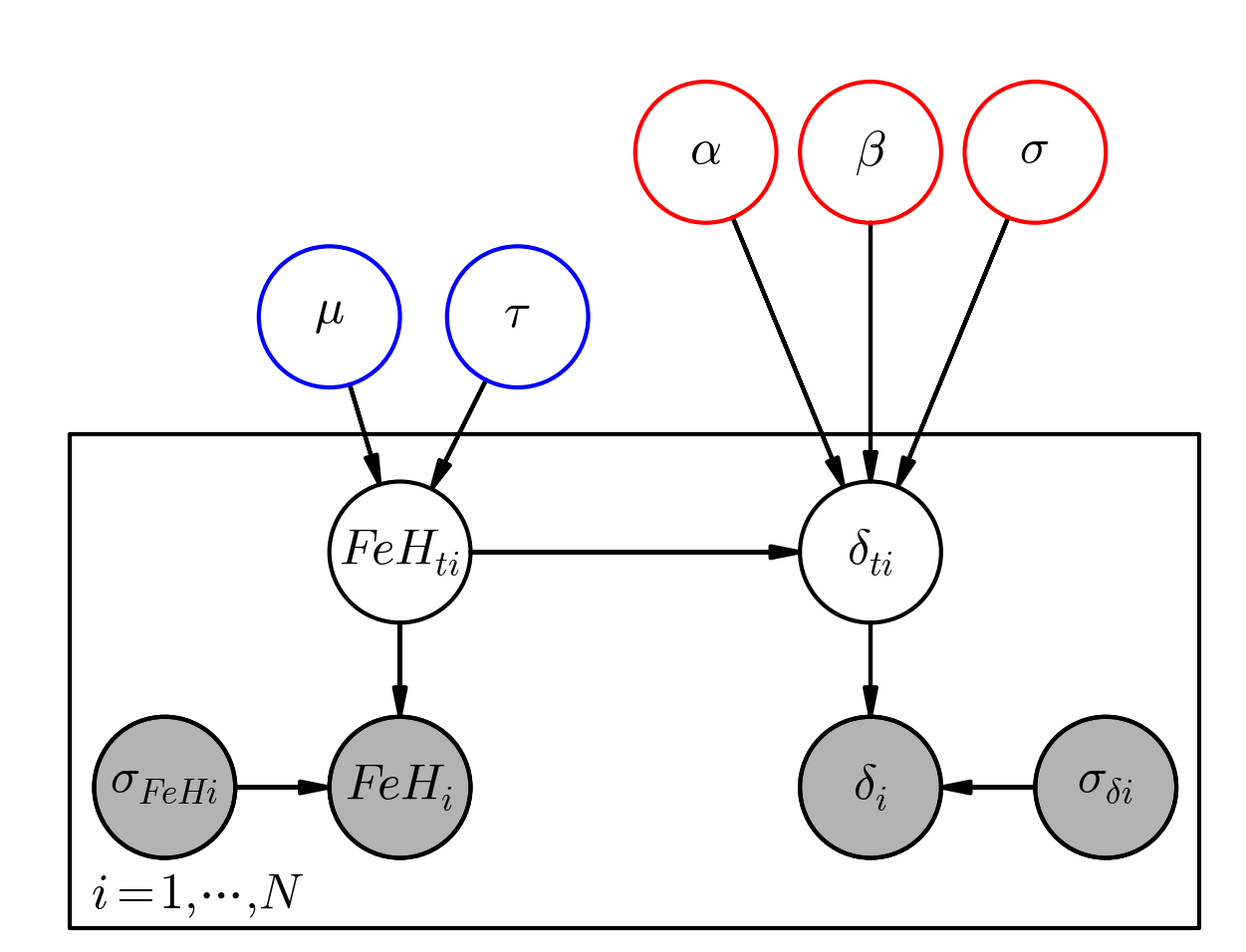}
\caption[test for the LoF] 
{A graphical model representing our HBM. The gray
nodes are the observed parameters. The true missing
parameters are in the white nodes. The blue nodes are the nuisance
parameters and the parameters of interest are in the red nodes.
The arrows indicate the conditional dependence. The definition of each parameter is:
\\\protect\begin{minipage}{\linewidth}
\begin{align*}
    FeH_i &= \text{ stellar metallicity of the i} ^\text{th} \text{ planet} \\
    \sigma_{FeHi} &= \text{uncertainty on the stellar metallicity of the i} ^\text{th} \text{ planet} \\
    \delta_i &= \text{ transit depth of the i} ^\text{th} \text{ planet} \\
    \sigma_{\delta i} &= \text{ uncertainty on the transit depth of the i} ^\text{th} \text{ planet} \\
    FeH_{ti} &= \text{ true stellar metallicity of the i} ^\text{th} \text{ planet} \\ 
    \delta_{ti} &= \text{ true transit depth of the i} ^\text{th} \text{ planet} \\    
    \mu \text{ and } \tau &= \text{nuissance parameters} \\
    \alpha \text{, } \beta \text{, and } \sigma &= \text{ parameters of the linear model} 
  \end{align*}	
  \protect\end{minipage}}
  \label{author1:fig1}
\end{figure}

Hierarchical Bayesian Modeling (hereafter HBM) 
allows us to derive the
uncertainties on the model parameters
and to relate the observed data to the
true unobserved data. 
Following \cite{Kelly:2007}, we constructed
the likelihood function in a simple way
to relate the parameters of interest to
the observed data taking into account
the measurement uncertainties. We
used this method to study the
correlation between the transit depth ($\delta$)
and the metallicity ($FeH$) of the host star. 
A graphical model of our hierarchical model is given in Fig.~\ref{author1:fig1}. 
Markov Chain Monte Carlo (hereafter MCMC) was performed using the
python package PySTAN\footnote{\url{http://mc-stan.org/}}, a package for Bayesian 
inference.
We ran the model with 4 Markov Chains, each of 5,000 iterations.
The first 50\% of each chain were
discarded as \q{burn-in} and the remaining samples were combined ending up
with 10,000 samples.

\section{Results}

\begin{figure}[t!]
 \centering
 \includegraphics[width=0.48\textwidth,clip]{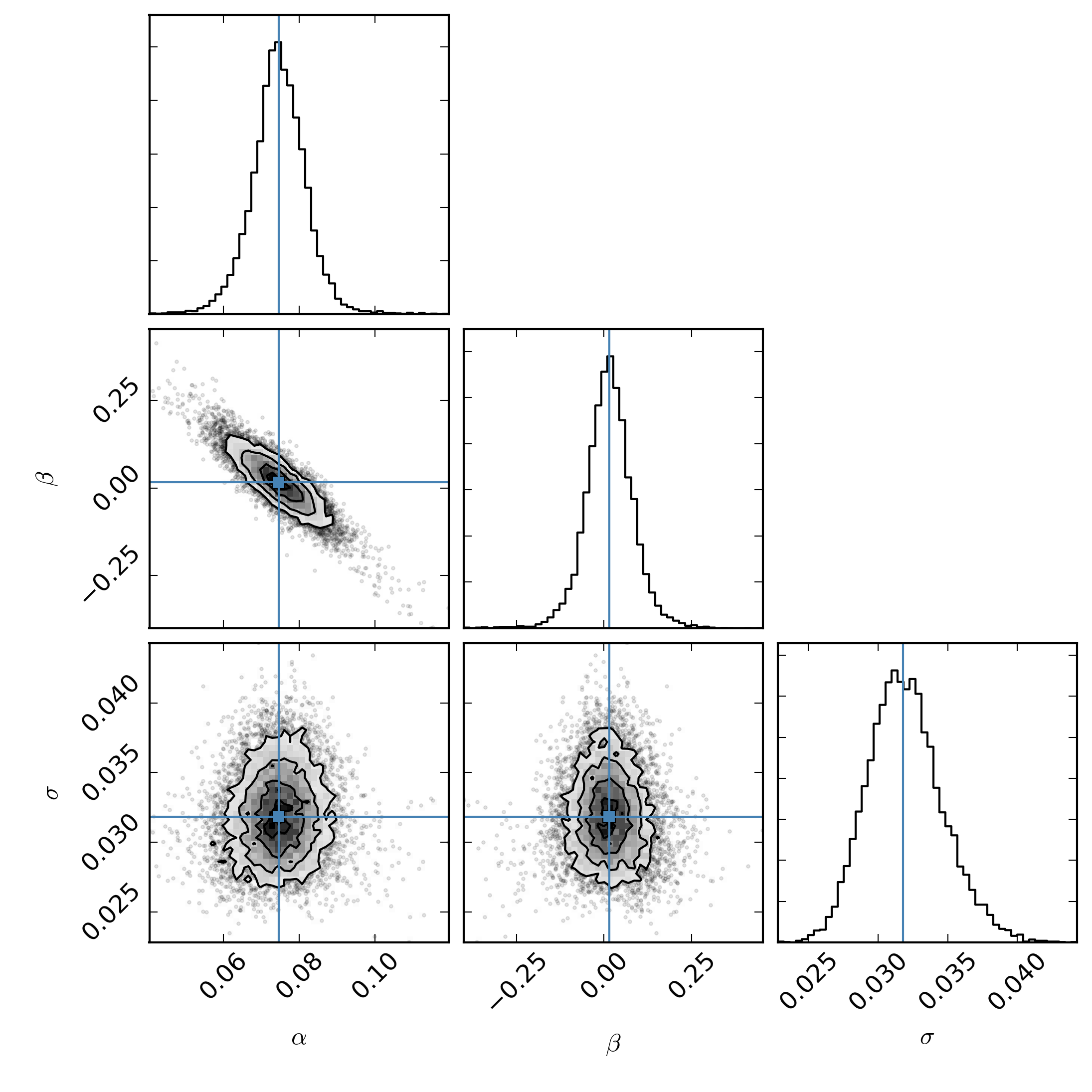}%
 \includegraphics[width=0.48\textwidth,clip]{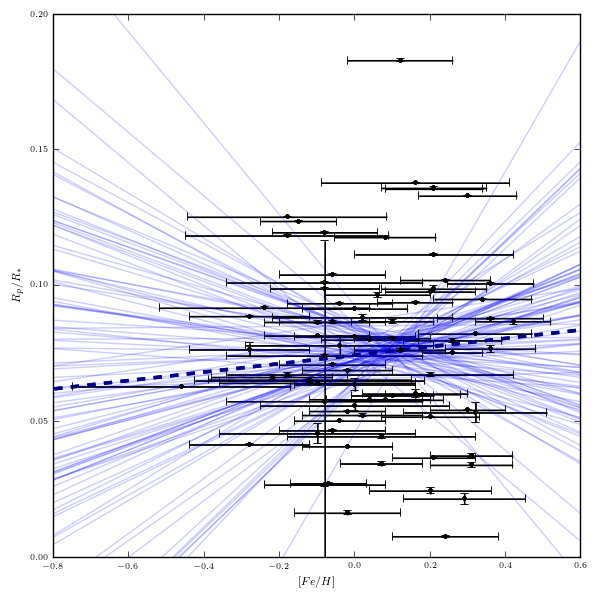}      
  \caption{{\bf Left:} Posterior probability distributions for the
parameters of the model \thetta\
as computed by MCMC, marginalized over the other parameters. 
{\bf Right:} Confirmed gas giant explanets and the best fit line. The light blue lines are
samples from the MCMC chain.}
  \label{author1:fig2}
\end{figure}

The posterior distributions for each of the parameters of interest (\thetta)
produced by running MCMC are shown in 
the left panel of Fig.~\ref{author1:fig2}. 
The equation of the \q{best-fit} linear model is 

\begin{equation}
\delta	= (0.07 \pm 0.014) + (0.02 \pm 0.14)FeH
\end{equation}

\noindent
with an intrinsic scatter of $\sigma = 0.03 \pm 0.005$. 
The transit depth of each confirmed exoplanet is 
plotted against the metallicity of the host star along with
their uncertainties in the right panel of Fig.~\ref{author1:fig2}.
The \q{best-fit} linear model (blue dotted line) is
shown along with 100 random samples from the MCMC chain (light blue).
It is clear that there
is a large intrinsic scatter which 
leads to the conclusion that there is no 
correlation between the transit depth and the stellar metallicity.
    
\section{Discussion}

In this work, we showed that there is no correlation between the transit 
depth of \kepler's giant exoplanets and the metallicity of the host star.
In particular, we demonstrated that there is a relatively large intrinsic
scatter in the relation. This result shows how crucial understanding those
discrepancies. Thus, they are directly related to the models of planetary
structure and formation.

For future work, this analysis should account for the selection effects 
and biases present in the \kepler\ survey. \cite{Gaidos:2013} 
reported the importance of including these effects in any statistical study.
The authors also showed that these selection effects lead to biases 
in the properties of transiting planets and their host stars, hence 
biasing the correlation.

\begin{acknowledgements}
This research has made use of the data collected by the \kepler\ mission and the NASA Exoplanet Archive, which is operated by the California Institute of Technology, under contract with the National Aeronautics and Space Administration under the Exoplanet Exploration Program.
\end{acknowledgements}

\bibliographystyle{aa}  
\bibliography{sf2a-template} 

\end{document}